\documentclass[sigconf]{acmart}

\usepackage{booktabs} 
\usepackage{epsfig,endnotes}
\usepackage{xspace}
\usepackage{acro}
\usepackage{pgf}
\usepackage{graphicx}
\usepackage{subfigure}
\usepackage{cleveref}
\usepackage{float}
\setcopyright{rightsretained}

\acmDOI{10.1145/3098243.3098246}

\acmISBN{978-1-4503-5084-6/17/07}

\acmConference{WiSec '17 }{July 18-20, 2017}{Boston, MA, USA}
\acmPrice{15.00}
\acmYear{2017} 
\copyrightyear{2017}

\begin{document}
\title{Using Hover to Compromise the Confidentiality of User Input on Android}


\author{Enis Ulqinaku}
\affiliation{%
  \institution{Sapienza University of Rome}
}
\email{ulqinaku@di.uniroma1.it}

\author{Luka Malisa}
\affiliation{%
  \institution{ETH Zurich, Switzerland}
}
\email{luka.malisa@inf.ethz.ch}

\author{Julinda Stefa}
\affiliation{%
  \institution{Sapienza University of Rome}
}
\email{stefa@di.uniroma1.it}

\author{Alessandro Mei}
\affiliation{%
  \institution{Sapienza University of Rome}
}
\email{mei@di.uniroma1.it}

\author{Srdjan {\v{C}}apkun}
\affiliation{%
  \institution{ETH Zurich, Switzerland}
}
\email{srdjan.capkun@inf.ethz.ch}

\begin{abstract}
We show that the new hover (floating touch) technology, available in a number of today's smartphone models, can be abused by malicious Android applications to record all touchscreen input into applications \emph{system-wide}. Leveraging this attack, a malicious application running on the system is able to capture sensitive input such as passwords and PINs, record all user's social interactions, as well as profile user's behavior. To evaluate our attack we implemented \emph{Hoover}, a proof-of-concept malicious application that runs in the background and records all input to all foreground applications. We evaluated Hoover with 20 users, across two different Android devices and two input methods, stylus and finger. In the case of touchscreen input by \emph{finger}, Hoover estimated the positions of users' clicks within an error of 100 pixels and keyboard input with an accuracy of 79\%. Hoover captured users' input by \emph{stylus} even more accurately, estimating users' clicks within 2 pixels and keyboard input with an accuracy of 98\%. Differently from existing well-known side channel attacks, this is the first work that proves the security implications of the hover technology and its potential to steal all user inputs with high granularity. We discuss ways of mitigating this attack and show that this cannot be done by simply restricting access to permissions or imposing additional cognitive load on the users since this would significantly constrain the intended use of the hover technology.

\copyrightyear{2017} 
\acmYear{2017} 
\setcopyright{acmcopyright}
\acmConference{WiSec '17 }{July 18-20, 2017}{Boston, MA, USA}\acmPrice{15.00}\acmDOI{10.1145/3098243.3098246}
\acmISBN{978-1-4503-5084-6/17/07}
\end{abstract}

%
%



\maketitle


\section{Introduction}
\label{sec:introduction}

The recent years witnessed a surge of \emph{input inference attacks}---attacks that infer (steal) either partial or all user input. This is not surprising, as these attacks can profile users and/or obtain sensitive user information such as login credentials, credit card numbers, personal correspondence, etc. Existing attacks are predominantly application-specific, and work by tricking the users into entering their information through phishing or UI redressing~\cite{embedded2013, taskJacking15, defenseSchemes2015, vigna15} (e.g., clickjacking~\cite{clickjackingBlackhat12}). Other attacks exploit readily available sensors on modern smartphones as side-channels. They infer user input based on readings of various sensors, such as the accelerometer~\cite{acCompliance2012}, gyroscope~\cite{gyroUSENIX14} and microphone~\cite{Narain2014}. Access to these sensors (microphone excluded) requires no special permissions on Android.

In this work, we introduce a novel user input inference attack for Android devices that is \emph{more accurate}, and more general than prior works. Our attack simultaneously affects all applications running on the device (it is \emph{system-wide}), and is not tailored for any given app. It enables continuous, precise collection of user input at a high granularity and is not sensitive to environmental conditions. The aforementioned approaches either focus on a particular input type (e.g., numerical keyboards), are application-specific, operate at a coarser granularity, and often only work under specific conditions (limited phone mobility, specific phone placement, limited environmental noise). Our attack is not based on a software vulnerability or system misconfiguration, but rather on a new and unexpected use of the emerging \emph{hover} (floating touch) technology. 

\begin{figure}[b]
        \centering
        \includegraphics[width=0.7\linewidth, keepaspectratio]{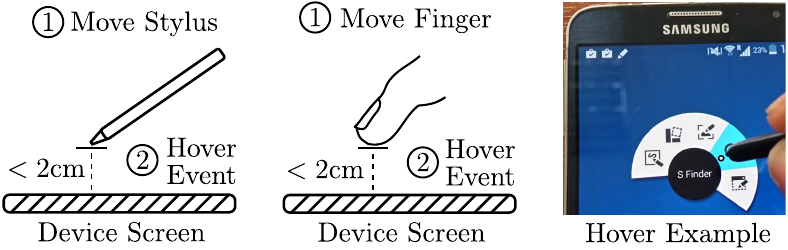}
        \caption{Hover technology. The input device creates special events (hover events) without touching the device screen. The rightmost part shows a user interacting with the phone without the input device touching the screen.}
        \label{fig:example_hover}
\end{figure}

The hover technology gained popularity when Samsung, one of the most prominent players in the mobile market, adopted it in its Galaxy S4, S5, and Note series. The attack presented in this work can therefore potentially affect millions of users~\cite{s4sales, s5sales, note2sales, note3sales}. The hover technology, illustrated in \Cref{fig:example_hover} produces a special type of event (hover events) that allow the user to interact with the device without physically touching its screen. We show how such hover events can be used to perform powerful, system-wide input inference attacks.

Our attack carefully creates and destroys overlay windows, right after each user tap to the foreground app, in order to capture just enough post-tap hover events to accurately infer the precise click coordinate on the screen. Previous phishing, clickjacking, and UI redressing techniques~\cite{embedded2013, taskJacking15, defenseSchemes2015, clickjackingBlackhat12} also create overlay windows, commonly using the {SYSTEM\_ALERT\_WINDOW} permission. Our attack does not rely on it: We present an implementation that does not require \emph{any permissions}. Furthermore, overlay windows in our case are exploited in a conceptually different manner. Our attack is continuous, completely transparent to the user, does not obstruct the user interaction with the foreground app, does not redirect the user to other malicious views, and does not deceive the user in any manner---a set of properties not offered by existing attacks. 

To evaluate our attack, we implemented \emph{Hoover}, a proof-of-concept malicious application that continuously runs in the background and records the hover input of all applications. However, to realize our attack we had to overcome technical challenges. Our initial experiments with the hover technology showed that hover events, unexpectedly, are predominantly not acquired directly over the point where the user clicked. Instead, the events were scattered over a wider area of the screen. Therefore, to successfully predict input event coordinates, we first needed to understand how users interact with smartphones. For this purpose we performed a user study with 20 participants interacting with one of two devices with \emph{Hoover} on it, in two different use-case scenarios: General clicking on the screen and typing regular English text. The hover events acquired by \emph{Hoover} were used to train a regression model to predict click coordinates, and a classifier to infer the keyboard keys typed.

We show that our attack works well in practice with both stylus and fingers as input devices. It infers general user finger taps with an error of 100px. In case of stylus as input device, the error is reduced to just 2px. Whereas, when applying the same adversary to the on-screen keyboard typing use-case, the accuracy of keyboard key inference results of 98\% and 79\% for stylus and finger, respectively.

A direct (and intuitive) implication of our attack is compromising the confidentiality of all user input, system-wide. For example, \emph{Hoover} can record various kinds of sensitive input, such as pins or passwords, as well as social interactions of the user (e.g., messaging apps, emails). However, there are also alternative, more subtle, implications. For example, Hoover could also profile the way the device owner interacts with the device, i.e., generate a biometric profile of the user. 
This profile could be used to, e.g., restrict the access only to the device owner, or to help an adversary bypass existing keystroke based biometric authentication mechanisms~\cite{iCanBeYou}.

We discuss possible countermeasures against our attack, and we observe that, what might seem as straightforward fixes, either cannot protect against the attack, or severely impact the usability of the system or of the hover technology.

To summarize, in this work we make the following contributions:
\begin{itemize}
        \item We introduce a novel and system-wide Android user-input inference attack, based on hover technology.
        \item We implement \emph{Hoover}, a proof-of-concept malicious app.
        \item We perform user studies, and show that \emph{Hoover} is accurate.
        \item We discuss possible countermeasures, and show that the attack is challenging to prevent.
\end{itemize}

The rest of this paper is organized as follows. In Section~\ref{sec:background} we describe background concepts regarding the hover technology and the view UI components in the Android OS. Section~\ref{sec:our_approach} states the problem considered in this work and describes our attack on a high-level. Successively, in Section~\ref{sec:evaluation} we present the implementation details of \emph{Hoover} and its evaluation. Our attack implications are then discussed in Section~\ref{sec:implications}, while Section~\ref{sec:countermeasures} presents the possible countermeasures. Section~\ref{sec:related} reviews related work in the area, and Section~\ref{sec:discussion} concludes the paper and outlines future work.

\section{Background}
\label{sec:background}

In this section we provide some background on the hover technology and the Alert Windows, a very common UI element used by many mobile apps in Android.

\subsection{Hover Events in Android}
The Hover (or floating touch) technology enables users to interact with mobile devices without physically touching the screen. We illustrate the concept in \Cref{fig:example_hover}. 
This technology was first introduced by the Sony Xperia Device~\cite{floatingtouch} in 2012, and is based on combining mutual capacitance and self-capacitance sensing. After the introduction by Sony, the hover technology was adopted by Asus in its Fonepad Note 6 device in late November 2013. It finally took over when Samsung, one of the biggest players in the market, used it in a series of devices including the Galaxy S4, S5, and the Galaxy Note~\cite{airviewS4}. Samsung alone has sold more than 100M devices supporting the hover technology~\cite{s4sales, s5sales, note2sales, note3sales}---all of them potential target of the attack described in this paper.

The hover is handled as follows: When the user interacts with the screen, the system is able to detect the position of the input device before touching it. In particular, when the input device is hovering within $20$mm from the screen (see Figure~\ref{fig:example_hover}), the operating system triggers a special type of user input event---\emph{the hover event}---at regular intervals. Apps that catch the event get the precise location of the input device over the screen in terms of $x$ and $y$ coordinates. Once the position of the input device is captured, it can then be dispatched to \emph{View Objects}---Android's building blocks for user interface---listening to the event. More in details, the flow of events generated by the OS while the user hovers and taps on the screen are as follows: When the input device gets close to the screen (less than $20$mm), the system starts firing a sequence of \emph{hover events} with the corresponding $(x, y)$ coordinates. A \emph{hover exit} event followed directly by a \emph{touch down} event are fired when the screen is touched. A \emph{touch up} event notifies the end of the touch. Afterwards, another series of \emph{hover} events are again fired as the user moves the input device away from the touching point. Finally, when the input device leaves the hovering area, i.e., is floating higher than 20mm from the screen, a \emph{hover exit} event is fired. 

\subsection{View Objects}
Android handles the visualization of system and app UI components on screen through the \emph{WindowManager} Interface~\cite{windowManager}. This is responsible for managing and generating the windows, views, buttons, images, and other floating objects on the screen. Depending on their purpose, the views can be generated so as to catch hover and touch events (active views, e.g., a button), or not (passive views, e.g., a mere image).  A given view's mode can be changed, however, from passive to active, and so on, by setting or unsetting specific flags through the \emph{updateViewLayout()} API of the \emph{WindowManager}  Interface. In particular, to make a view passive, one has to set the \emph{FLAG{$\_$}NOT{$\_$}FOCUSABLE} and \emph{FLAG{$\_$}NOT{$\_$}TOUCHABLE}. The first flag avoids that the view gets key input focus. The second flag disables the ability to intercept touch or hover events. These two flags make so that a static view does not interfere with the normal usage of the device, even in the case when it is on top of all other windows. In addition, a given view can learn precisely when, somewhere on screen and outside the view, a click was issued, without knowing the position of the click. This is made possible by setting the \emph{FLAG{$\_$}WATCH{$\_$}OUTSIDE{$\_$}TOUCH} of the view.

In our work we use views that are on top of all the other objects, including the views of the foreground app. These particular views can be implemented either as Alert Windows or as Toast Windows~\cite{toast}. Alert Windows are used by off-the-shelf apps like Text Messaging or Phone and by many other apps---a search of the Play market through the IzzyOnDroid online crawler~\cite{storecrawler} reveals that there are more than 600 apps with hundreds of millions of downloads that use Alert Windows.
To generate Alert Windows the \emph{WindowManager} interface uses the \emph{SYSTEM{$\_$}ALERT{$\_$}WINDOW} permission, that must be held by the service that creates the view. 
However, the functionalities that we need for our attack can be implemented without requiring any permission at all by using the \emph{Toast} class. This implementation is more complex, due to technicalities of Toast Windows that are trickier to handle, therefore we proceed by describing our attack with Alert Windows and later, in Section~\ref{sec:independent}, we show how to get an implementation of our attack requiring no particular permission.

\section{Our Attack}
\label{sec:our_approach}

The goal of our attack is to track every click the user makes with both high precision (e.g., low estimation error) and high granularity (e.g., at the level of pressed keyboard keys). The attack should work with either finger or stylus as input device, while the user is interacting with a device that supports the hover feature. Furthermore, the attack should not be detected by the user, i.e.,  the attack should not obstruct normal user interaction with the device in any way. 

Before describing our attack, we state our assumptions and adversarial model.

\subsection{Assumptions and Adversarial Model}
\label{sec:system_model}
We assume the user is operating a mobile device that supports the hover technology. The user can interact with the mobile with either a stylus, or a single finger, without any restrictions. 

We consider the scenario where the attacker controls a malicious app installed on the user device. The goal is to violate the confidentiality of user input without being detected. In our first, easier to describe implementation, the malware has access to two permissions only: The \emph{SYSTEM$\_$ALERT$\_$WINDOW}, a permission common in popular apps as discussed in the previous section, and the \emph{INTERNET} permission---so widespread that Android designated it as a \emph{PROTECTION$\_$NORMAL} protection level~\cite{androidPermissions}. This indicates that it is not harmful and is granted to all apps that require it without asking the user. Then, we describe a way to remove the \emph{SYSTEM$\_$ALERT$\_$WINDOW} permission with an alternative, more complex implementation that uses no particular permission.

\subsection{Attack Overview}
\label{sec:attack_overview}
To track the input device in-between clicks we exploit the way Android OS delivers hover events to apps. When a user clicks on the screen, the following sequence of events with coordinates and time stamps is generated (see Section~\ref{sec:background}): \textit{hover(s)} (input device floating); \emph{hover exit} and \emph{touch down} (on click); \emph{touch up} (end of click); \textit{hover(s)} (input device floating again).

To observe these events, a malicious app can generate a transparent Alert Window overlay if it holds the \emph{SYSTEM$\_$ALERT$\_$WINDOW} permission, otherwise it can use the \emph{Toast} class to create the overlay and implement the attack as described in Section~\ref{sec:independent}.
Recall that the Alert Window components are placed on top of any other view by the Android system (see Section~\ref{sec:background}). Once created, the overlay could catch the sequence of hover events fired during clicks and would be able to track the input device. However, doing so in a stealthy way, without obstructing the interaction of the user
with the actual apps, is not trivial. The reason is that Android sends \emph{hover events} only to those views that receive \emph{touch events}. In addition, the system limits the ``consumption'' of a touch stream, all events in between including \emph{touch down} and \emph{touch up} to one view only. So, a malicious overlay tracking the input device would either catch both hovering \emph{hover events} and the touch, thus impeding the touch to go to the real app, or none of them, thus impeding the malware to infer the user input.

\subsection{Achieving Stealthiness}
\label{sec:stealthiness}
The malicious app controlled by the adversary cannot directly and stealthily observe click events. We show that, instead, it can infer the clicks stealthily by observing hover events preceding and following user clicks. By doing so accurately, the adversary will be able to infer the user input without interfering  with user interaction. 

In more details, our attack is constructed as follows: The malicious app generates a fully-transparent Alert Window overlay which covers the entire screen. The overlay is placed by the system on top of any other window view, including that of the app that the user is using. Therefore, the malware, thanks to the overlay, can track the hover events. However, the malicious view should go from \emph{active} (catch all events) to \emph{passive} (let them pass to the underneath app) in a ``smart way'' in time, so that the touch events go to the real app while the hovering coordinates are caught by the malware. The malware achieves this by creating and removing the malicious overlay appropriately, through the WindowManager APIs, in a way that it does not interfere with the user interaction. This procedure is detailed in the next section.

\subsection{Catching Click and Hover Events}
\label{sec:attack_get_hovers}
We implement our adversary (malware) as a background service, always up and running on the victim device. That said, the main challenge of the malware is to know the exact time when to switch the overlay from \emph{active} (add it on screen) to \emph{passive} mode (remove it), and back to \emph{active} mode again. Note that, to guarantee the attack stealthiness, we can catch hover events only. Not the ones that regard the actual touch, which should go to the app the user is interacting with. Therefore, foreseeing when the user is going to stop hovering the input device in order to actually click on the screen is not simple. We approach the issue in the following way: Through \emph{WindowManager} the malware actually makes use of two views. One is the fully transparent Alert Window overlay mentioned earlier in this section. The second view, which we call \emph{Listener} and has a size of 0px, does not catch neither hover coordinates nor clicks. Its purpose is to let the malware know when a click happens, only. The Hoover malware will then use this information to remove/re-create the transparent overlay.

\subsubsection{Inferring Click Times}
All user clicks happen outside the Listener view---it has a size of 0px. In addition, this view has the \emph{FLAG{$\_$}WATCH{$\_$}OUTSIDE{$\_$}TOUCH} set, so it is notified when the \emph{touch down} event 
corresponding to the click is fired. As a result, the malware infers the timestamp of the click, though it cannot know the position on the screen (see Step 1 in Figure~\ref{fig:overview}). 

\subsubsection{Catching Post-click Hover Events}
\begin{figure}[t]
        \centering
        \includegraphics[ width=0.6\linewidth, keepaspectratio]{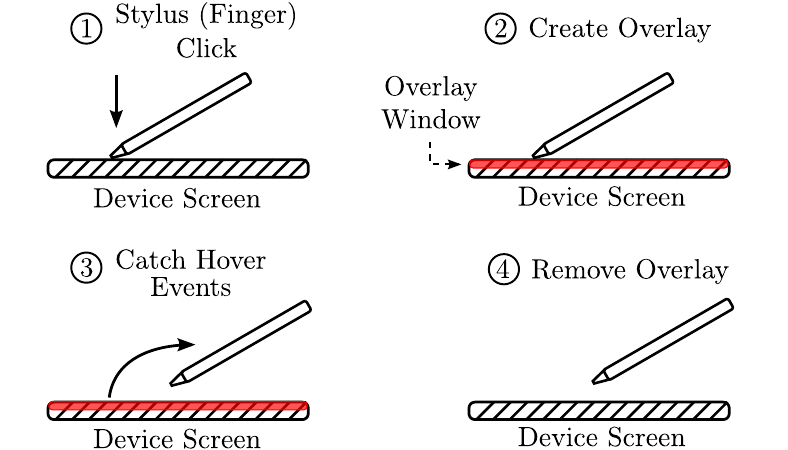}
        \caption{Hoover catching post-click hover events with the transparent malicious overlay.}
        \label{fig:overview}
\end{figure}

In order to infer the click position, the attack activates a whole-screen transparent overlay right after the touch down event is fired and the click is delivered to the legitimate application (see Step 2 in Figure~\ref{fig:overview}). This guarantees that the attack does not interfere with the normal usability of the device. The overlay, from that moment on, intercepts the hover events fired as the input device moves away from the position of the click towards the position of the next click (see Step 3 in Figure~\ref{fig:overview}).

Differently from the Listener view, which cannot interfere with the user-device interaction because of its size of 0px, the overlay cannot be always active (present on the screen). Otherwise it will obstruct the next clicks of the user intended for the app she is using. At the same time, the overlay must remain active long enough to capture a number of hover events following the click sufficient to perform an accurate click location inference. Our experiments show that, with the devices considered in this work, hover events are fired every 19ms in average by the system. In addition, we find that 70ms of activation time is a good trade-off between catching enough hover events for click inference and not interfering with the user-device interaction. 
This includes additional usability features of apps, different from actual clicks, like the visualization of hint words when the finger is above a button while typing on the keyboard. After the activation time elapses, the overlay is removed again (see Step 4 in Figure~\ref{fig:overview}).

\subsection{Inferring Click Positions}
\label{sec:attack_machine_learning}

\begin{figure}[t]
        \centering
        \includegraphics[width=0.6\linewidth, keepaspectratio]{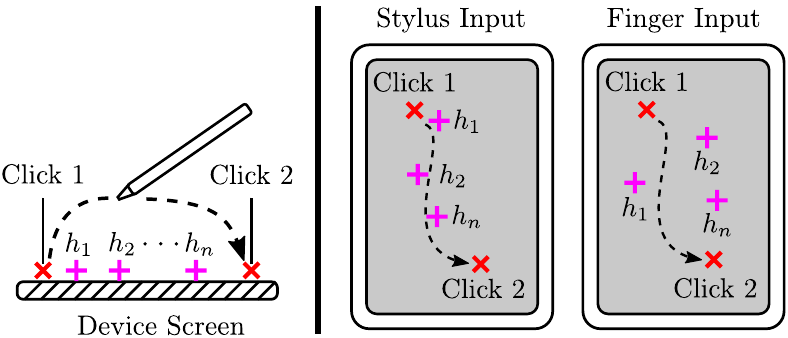}
        \caption{Example of hover events collected by Hoover. In case of stylus input, hover events ($h_1, h_2, \ldots, h_n$) follow quite faithfully the stylus path, but they are scattered over a wider area in case of finger.}
        \label{fig:hovers}
\end{figure}

At this stage, the malware has collected a set of post-click hover events for each user click. Starting from the information collected, the goal of the attacker is to infer the position of each user click as accurately as possible. A solution could be to determine the click position based on the position of the first post-click hover event only. While this approach works well with stylus clicks, it is not good enough to determine finger clicks. The reason is that the stylus, having a smaller pointing surface, generates hover events which tend to follow the trajectory of user movement (see Figure~\ref{fig:hovers}). As a result, the first post-click hover event (respectively, the last hover before the click) tend to be very close to the position of the corresponding click. Conversely, the surface of the finger is considerably larger than that of the stylus pointer. Therefore, hover events, including post-click ones, do not follow that precisely the trajectory of the movement as in the stylus case. This is confirmed by our initial experiment results that show that the position of the first post-click hover captured is rarely strictly over the position of the click itself.

For this reason, in order to improve the accuracy of click inference of our approach we decided to employ machine learning tools that consider not only the first post-click hover event, but all those captured in the 70ms of the activation of the overlay. In particular, for the general input-inference attack we employ a \emph{regression model}. For the keyboard-related attacks (key inference) we make use of a classifier. On a high level, given the set of post-click captured hover events ($h_1, h_2, \ldots, h_n$), a regression model answers the question: \emph{``Which is the screen position clicked by the user?''}. Similarly, the classifier outputs the key that was most likely pressed by the user. To evaluate our attack we experimented with various regression and classifier models implemented within the analyzer component of the attack using the scikit-learn~\cite{scikit-learn} framework. We report on the result in the next section. 

In our initial experiments, we noticed that different users exhibit different hover event patterns. Some users move the input devices faster than others. In the case of fingers, the shape and size of the users' hands resulted in significantly different hover patterns. To achieve accurate and robust click predictions, we need to train our regression and classifier models with data from a variety of users. For that purpose, we performed two user studies that we describe in the next section.

\section{Evaluation}
\label{sec:evaluation}
\subsection{The Attack (Malware) Prototype and Experimental Setup}
\label{sec:setup}
To evaluate the attack presented in this work we implemented a prototype for the Android OS called the \emph{Hoover}. The prototype operates in two logically separated steps: It first collects hover events (as described in Section~\ref{sec:our_approach}) and then it analyzes them to predict user click coordinates on screen. We implemented the two steps as two distinct components. Both components could easily run simultaneously on the user device. However, in our experiments we opted for their functional split, as it facilitates our analysis: The hover collecting component was implemented as a malicious Android app and runs on the user device. The analyzer was implemented in Python and runs on our remote server. The communication among the two is made possible through the \emph{INTERNET} permission held by the malicious app, a standard permission that Android now grants by default to all apps requesting it, without user intervention.

Uploading collected hover events on the remote server does not incur a high bandwidth cost. For example, we actively used a device for 4 hours, during which our malicious app collected events. The malware collected hover events for approximately 3,800 user clicks. The size of the encoded hover event data is 40 Bytes per click and the total data to be uploaded amounts to a modest 150kB. We obtained this data during a heavy usage of the device and the numbers represent an upper bound. So, we believe that, in a real-life usage scenario, the average amount of clicks collected by a standard user will be significantly less.

Finally, for the experiments we recruited 20 participants, whose demography is detailed in the next section. The evaluation of Hoover was done in two different attack scenarios: A general one, in which we assume the user is clicking anywhere in the screen and a more specific one, targeting on-screen keyboard input of regular text. We performed a large number of experiments with both input methods, the stylus and the finger, and on two different devices whose specifics are shown in Table~\ref{tab:device_specs}. However, the ideas and insights on which Hoover operates are generic and do not rely on any particularity of the devices. Therefore, we believe that it will work just as well on other hover-supporting Android devices.

\begin{table}
	\begin{center}
	\resizebox{\columnwidth}{!}{%
		\begin{tabular}{l  l  l}
			\toprule
			\textbf{Device Type} & \textbf{Operating System} & \textbf{Input Method}\\ 
			\midrule
			Samsung Galaxy S5 & Cyanogenmod 12.1 & Finger\\
			Samsung Galaxy Note 3 Neo & Android 4.4.2 & Stylus\\
			\bottomrule
		\end{tabular}
	}
	\end{center}
	\caption{Specifics of the devices used in the experiments.} 
	\label{tab:device_specs}
\end{table}

\subsection{Use-cases and Participant Recruitment}
\label{sec:user_study}

In this section we describe each use-case scenario in detail, and report on the participants recruited for the evaluation of our attack. 
\textbf{Use-case I (Generic clicks)}. The goal of the first use-case scenario was to collect information on user clicks anywhere on the screen. For this, the users were asked to play a custom game: They had to recurrently click on a ball shown on random positions on the screen after each click. This use-case scenario lasted 2 minutes.

\textbf{Use-case II (Regular text)}. The second use-case scenario targeted on-screen keyboard input. The participants were instructed to type a paragraph from George Orwell's ``1984'' book. Each paragraph contained, on average, 250 characters of text in the English language, including punctuation marks.

Each use-case scenario was repeated 3 times by the participants. In the first iteration they used their thumb as input device. In the second iteration they used their index finger, whereas in the third and last one, the stylus. During each use-case and corresponding iterations we recorded all user click coordinates and hover events that followed them.

\subsubsection{Participant Recruitment} For the experiments we enrolled a total of 20 volunteers from a university campus. We present the demographic details of our participants in Table~\ref{tab:demographics}. 
The users operated on the devices of our testbed (Table~\ref{tab:device_specs}) with the Hoover malware running in the background. Our set of participants (Table~\ref{tab:demographics}) includes mainly younger population whose input will typically be faster; we therefore believe that the Hoover accuracy might only improve in the more general population. We plan to evaluate this in more detail as a part of our future work. 

As a result of our on-field experiments with the 20 participants, we collected approximately 24,000 user clicks. Furthermore, the malware collected hover events for 70ms following each click. Around 17 K clicks were of various keyboard keys, while the remaining 7 K were collected from users playing the ball game. Users did not observe lagging or other signs of an ongoing attack during experiments.

\textbf{Ethical considerations.} The experiments were carried out by lending to each of the volunteers our own customized devices. 
At no point did we require participants to use their own devices or provide any private or sensitive information like usernames or passwords. Consequently, and accordingly to the policy of our IRB, we didn't need any explicit authorization to perform our experiments.

\begin{table}
	\centering
	\resizebox{\columnwidth}{!}{%
		\begin{tabular}{l  c  c | c  c  c | c  c  c | c}
		\toprule
  		&  \multicolumn{2}{c |}{\textbf{Gender}} & \multicolumn{3}{c |}{\textbf{Education}} & 	\multicolumn{3}{c |}{\textbf{Age}} & {\textbf{Total}}\\
		  & M & F & BSc & MSc & PhD & 20-25 & 25-30 & 30-35 & {}\\
		\midrule
		Participants  & 15 & 5 & 3 & 5 & 12 & 7 & 7 & 6 & 20\\
		\bottomrule
		\end{tabular}
	}
	\caption{Demographics of experiment participants.}
	\label{tab:demographics}
\end{table}

\subsection{Post-click Hover Collection Duration}
\label{sec:collection_duration}
A first aspect to investigate is for how long Hoover should keep the malicious overlay active without obstructing the next click of the user. The results showed that in 95\% of the cases, the inter-click time (interval among two consecutive clicks) is larger than 180ms. 

We then investigated how the number of post-click hover events impacts the prediction accuracy. For this reason, we performed a preliminary experimental study with just two participants. The initial results showed that the accuracy increases proportionally to the number of hover events considered. However, after the first 4 events, the accuracy gain is less than 1\% (78\% for 4 events, and 79\% for 5 events). Therefore, for the evaluation of the Hoover prototype we choose to use only 4 post-click hover events. This choice impacted the time that Hoover keeps the malicious overlay active for, i.e., its post-click hover event collection time. Indeed, we observed that 70ms were more than enough, as the first 4 post-click hover events were always fired within 70ms after the user click.

Lastly, note that our choice of 70ms is quite conservative when compared with the 180ms of inter-click time observed in our experiments. However, as we will see in the next sections, the prediction results with the Hoover prototype are quite high. On the one hand, a longer collection time would increase the number of post-hover events captured which could improve the accuracy of the malware in inferring user input. On the other hand, a static, longer collection time risks to expose the adversary to users whose click speed is very high---higher than those of the users in our experiment. That said, a more sophisticated adversary could start off with an arbitrarily short collection window and dynamically adapt it to the victim's typing speed.

\begin{figure*}
\centering
        \subfigure[Click position RMSE in pixels.]{
    		\includegraphics[width=0.22\linewidth]{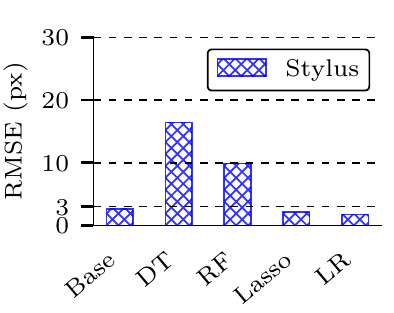}
		\label{fig:regressor_stylus}}\quad
 	 \subfigure[Click position RMSE in pixels.]{
  		\includegraphics[width=0.22\linewidth]{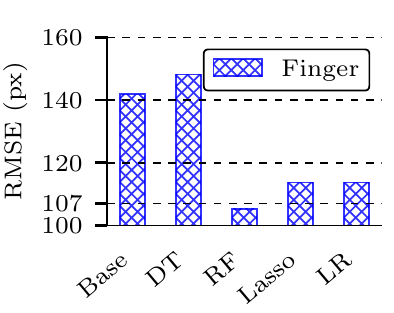}
		\label{fig:regressor_finger}}\quad
    \subfigure[Key prediction accuracy. Standard dev $\leq1$\% with all models.]{
    		\includegraphics[width=0.22\linewidth]{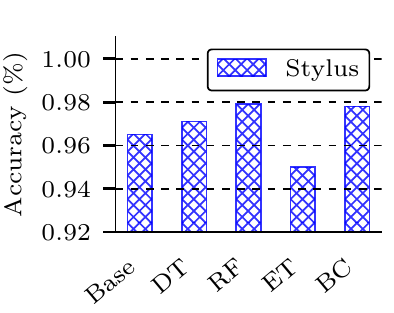}
		\label{fig:classifier_stylus}}\quad
 	 \subfigure[Key prediction accuracy. Standard dev $\leq1$\% with all models]{
  		\includegraphics[width=0.22\linewidth]{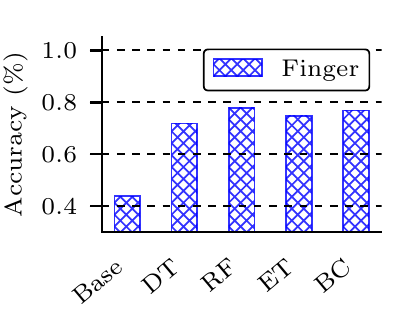}
		\label{fig:classifier_finger}}
\caption{Evaluation for Use-case I (figures~\ref{fig:regressor_stylus} and~\ref{fig:regressor_finger}) and Use-case II (figures~\ref{fig:classifier_stylus} and~\ref{fig:classifier_finger}). \textbf{Base}: Baseline, \textbf{DT}: Decision Tree, \textbf{RF}: Random Forest, \textbf{Lasso}: lasso regression, \textbf{LR}: linear regression, \textbf{ET}: extra trees classifier, \textbf{BC}: bagging classifier.}
\label{fig:prediction_results}
\end{figure*}


\subsection{Hoover Accuracy in Click Inference}
\label{sec:estimators}
Here we present the experimental results regarding the effectiveness and precision of Hoover to infer the coordinates of user clicks. Once Hoover obtains the post-click hover events from the user, it sends them to the machine-learning based analyzer running on the remote server (see Section~\ref{sec:attack_machine_learning}).


\subsubsection{Inferring Coordinates of General User Clicks}
The analyzer employs a regression model to infer the user click position on screen. Intuitively, the accuracy of the results depends on the model used for the prediction. Therefore, we experimented with a number of different models. In particular, we used two linear models (Lasso and linear regression), a decision tree, and an ensemble learning method (random forests)~\cite{scikit-learn}. The input to each model were the $(x, y)$ coordinates of the post-click hover events captured by Hoover (see Section~\ref{sec:our_approach}) for every user and click. The output consists of the coordinates of the predicted click position. As a benchmark baseline, we exploit a straightforward strategy that outputs the coordinates of the first post-click hover event observed. 

We used the leave-one-out cross-validation; i.e, for every user click validated the training was done on all other samples (user clicks). The prediction result for all click samples in our dataset obtained with the 20 participants in the experiment are presented in Figures~\ref{fig:regressor_stylus} and~\ref{fig:regressor_finger} for respectively the stylus and the finger. We see that the various regression models perform differently, in terms of Root Mean Square Error (RMSE).

First, we observe that, for all regression models, the finger-related results are less accurate than the stylus related ones. This is expected, as the hover detection technology is more accurate with the stylus (the hover events follow its movement more faithfully) than with the finger (its hover events are more scattered over the phone's screen). 
Nonetheless, in both cases the prediction works quite well. In particular, the estimation error with the stylus drops down to just 2 pixels. Consider that the screen size of the Note 3 Neo, the smallest device used in the experiments, is of $720 \times 1280$px. 

Lastly, we note that in the stylus case (see Figure~\ref{fig:regressor_stylus}) simple linear models perform better than more complex ones. This is not the case when the finger is used as an input device (see Figure~\ref{fig:regressor_finger}). Indeed, in this case the best predictions are given by the complex Random Forest model, followed by the linear regression. We believe that this is again due to the highest precision with which stylus hovers are captured by the screen w.r.t. those issued by the finger.

\subsubsection{Inferring On-Screen Keyboard Input}
To infer the keys typed by the users in the keyboard-based use-case we could follow a straightforward approach: (1) infer the corresponding click coordinates with the previous methodology, (2) observe that the predicted click coordinates fall within some key's area, and (3) output that key as the prediction result. 

As discussed in the previous section, the click prediction in the stylus case and with the linear regression model results being very accurate---only a 2px error within the actual click coordinate. So, the above straightforward solution works well for the stylus. However, the procedure is ill-suited for the finger case, where the error to predict the coordinates of the clicks is considerably larger (see Figure~\ref{fig:prediction_results}). For this reason, we take an alternative approach and pose the question as the following classification problem: ``\textit{Given the post-click hover events observed, which is the keyboard key pressed by the user?}''. Again, we experiment with several classification models: Two based on trees (decision trees and extra trees), the Bagging Classifier, and the random forest approach~\cite{scikit-learn}. Similarly to the regression case, we use a baseline model as a benchmark. The baseline simply transforms the coordinates of the first post-click hover event into the key whose area covers the coordinate. The results were obtained using 10-fold cross-validation.

The results of inferring regular text---Use-case II---are shown in Figure~\ref{fig:classifier_stylus} and~\ref{fig:classifier_finger} for respectively the stylus and the finger. First, we observe that the random forest (RF) method is the most accurate in key-prediction for both input methods---79\% for the finger (see Figure~\ref{fig:classifier_finger}) and up to 98\% for the stylus (see Figure~\ref{fig:classifier_stylus}). It is worth observing that in the finger case, the performance gap between the baseline and the more complex random forest approach significantly increases: It passes from 40\% (baseline) to 79\% (random forest) (see Figure~\ref{fig:classifier_finger}). Meanwhile, with the stylus, all of the approaches yield accurate results. In particular, the straightforward baseline approach is just 1\% away from the 98\% of accuracy achieved by the best performing random forest method (see Figure~\ref{fig:classifier_stylus}).

These results show that Hoover is quite accurate in stealing the text typed by the user with the on-screen keyboard. In addition, we believe that the accuracy could be improved by applying more complex dictionary-based corrections.

\subsection{Distinguish Keyboard Input from Other Clicks}
\label{sec:keyboard_detection}
Hoover collects all kind of user clicks. So, it needs to differentiate among on-screen keyboard taps and other types of clicks. One possible way is through side channels, e.g., \emph{/proc} folder.
Hoover could employ techniques similar to~\cite{activityAttack} to understand when the user is typing. However, we cannot just rely on the approach that uses the \emph{/proc} folder for the keyboard detection for two reasons. First, it is not fully accurate~\cite{activityAttack}, and it presents both false positives and negatives. Second, we cannot be sure that the \emph{/proc} folder will always be available and freely accessible for all apps.

We, therefore, implement a simple heuristic for this problem. The heuristic exploits the fact that the on-screen keyboard is shown at the bottom of the device's screen. Therefore, when a user is typing, the clicks are mostly directed towards the screen area covered by the keyboard. A straightforward methodology is to employ an estimator  to distinguish, among all user clicks, those that target keyboard keys. This solution would never present false negatives. However, it could result in some false positives. Indeed, a user could click on the lower part of the screen for many purposes: While playing a game that involves clicks, to start an app whose icon is located in that area, and so on.

To filter out clicks that could yield false positives we further refine our heuristic. The idea is simple: If the user is actually typing, she will issue a large number of consecutive clicks on the lower part of the screen. So, we filter out particularly short click sequences (less than 4 chars) that are unlikely to be even usernames or passwords. In addition, we empirically observed that, after the user clicks on a textbox to start typing, at least 500ms elapse till she types the first key. This is the time needed by the keyboard service to load it on the screen. We added the corresponding condition to our heuristic to further reduce the false positives.

We gathered data for 48 hours from a phone in normal usage (e.g., chatting, browsing, calling) to evaluate the heuristics. We collected clicks, corresponding hover events, and timestamps of the moments when the user starts (and stops) interacting with the keyboard. The false negative rate is 0 for both heuristics. The simple version has a false positive rate of 14.1\%. The refined version drops it down to 10.76\% (a 24\% improvement).

We implemented the heuristics only as a proof-of-concept. We believe that a more sophisticated refinement that also includes the difference between typing and clicking touch times (for how long the user leans the input device on screen during clicks) could considerably improve the false positive rate. However, these improvements are out of the scope of this work.

\subsection{Making Hoover Independent from \texttt{SYSTEM\_ALERT\_WINDOW}}
\label{sec:independent}
So far we described the Hoover implementation through Alert Windows, which require the \emph{SYSTEM\_ALERT\_WINDOW} permission. Here, we show how we can achieve the same functionalities in an alternative, permission free, though slightly more complex way: Through \emph{Toast} class~\cite{toast}.

The \emph{Toast} class allows to generate notifications or quick messages regarding some aspect of the system.  An example is the window that shows the volume control while the user is raising up or down the volume. They do not require any specific permission and can be employed by any service or user app. Most importantly, just like Alert Windows, Toasts can capture hover events, can contain fully customized objects of the View class, and are always shown on top of any other window, including the foreground app. Therefore, both the Listener and the transparent overlay can be generated as Toast Windows.
 
As a proof of concept we implemented a version of Hoover with Toast Windows. Android limits the activity time of a Toast to just a few seconds. This makes it trickier to implement the Listener view, which is supposed to stay always on screen. In fact, with the \emph{Toast} class, Hoover periodically calls the  \emph{toast.show()} method on the Listener before its expiration. The problem does not sustain with the transparent overlay, which is only shown for 70ms after each click detected by the Listener, as we already discussed in the previous sections. After the stream of hover events are collected we remove the overlay and activate the Listener again. In this way we implement Hoover with the same functionalities but without any particular permission.

\subsection{Further Attack Improvements}
The results previously discussed show that hover events can be used to accurately infer user input, be it general click positions or keyboard keys. 

In this section we list two additional techniques that, in our belief, could improve the attack and its accuracy. 

\textbf{Language model.} In our evaluation, we considered the worst case scenario, where the attacker does not make any assumptions on the language of the input text. Although the text typed by the users in the experiments was in English, it could have been in any arbitrary language. 
A more sophisticated attacker could first detect the language the user is typing in. Then, after the key inferring methods we described, apply additional error correction algorithms to improve the accuracy. 

\textbf{Per-user model.} In our evaluation both the regression models and classifiers were trained on data obtained from all users. That is, for each strategy we created a single regression and classification model that was then used to evaluate all users. 
This model has the advantage that, if a new user is attacked, the attack starts working right away with the results we show in the experiments. However, it is reasonable to think that a per-user model could result in a considerably higher accuracy. We could not fully verify this intuition on our dataset as we did not have sufficient per-user data for all participants. However, we did a preliminary evaluation on the two users with the most data points. The result with separate per-user model training showed a considerable improvement, particularly with the finger typed input. Indeed, The accuracy of keyboard key inference increased from 79\% (all users) to 83\% for the first user and 86\% for the second one.

\subsection{Alternative Keyboard Input Methods}
Our attack is very effective and accurate with typing. However, it cannot be as effective with swiping text. Indeed, Hoover infers coordinates only after the input device leaves the screen. With swiping, this translates in inferring only the last character of each word swiped by the user. That said, it is important to note that swiping is not enabled for e.g., password-like fields and characters such as numbers or symbols, which need to be typed and not swiped by the user. Therefore, even with swiping, Hoover would still be effective in stealing user sensitive information like passwords or pin numbers.

In our attack we assume that a regular keyboard is used. However, users could employ complex security mechanisms that, e.g., customize keyboards and rearrange the keys each time a user writes or enters her credentials. These types of mechanisms would certainly mitigate our attack: Hoover would not be able to map coordinates to keys correctly. However, at the same time the usability of the device would considerably decrease as the users would find it very difficult to write on a keyboard whose keys are rearranged each time. Consequently, it is very likely that systems would tend to restrict the protection mechanism to very sensitive information like PIN numbers and credentials, leaving texts, emails, and other types of messaging sequences still vulnerable to our attack.

\section{Implications of the attack}
\label{sec:implications}

The output of our attack is a stream of user clicks inferred by Hoover with corresponding timestamps. In the on-screen keyboard input use-case scenario, the output stream is converted into keyboard keys that the user has typed. 
In this section we discuss possible implications of the attack, techniques and ideas exploited therein. 

\subsection{Violation of User Privacy} 
A first and direct implication of our attack is the violation of user privacy. Indeed, a more in-depth analysis of the stream of clicks could reveal a lot of sensitive information regarding the device owner. To see why, consider the following output of our attack:
\begin{quote}
	\emph{john doe<CLICK>hey hohn, tomorrow at noon, downtown starbucks is fine with me.<CLICK> <CLICK>google.com <CLICK>paypal<CLICK>jane.doe<CLICK>hane1984}
\end{quote}

At first glance we quickly understand that the first part of the corresponding user click operations were to either send an email or a text message. Not only that, we also understand who is the recipient of the message---probably John---that the user is meeting the next day, and we uncover the place and the time of the meeting. Similarly, the second part of the sequence shows that the user googled the word \emph{paypal} to find a link to the website, that she most probably logged in it afterwards, that her name is Jane Doe and that her credentials of her \emph{Paypal} account are probably \emph{jane.doe} (username) and \emph{jane1984} (password). This is just a simple example that shows how easily Hoover, starting from just a stream of user clicks, can infer very sensitive information about a user. Intuitively, it also shows that if an adversary obtains all user input, finding passwords among other text is much easier than random guessing.

Another thing to observe in the above example is that the output contains errors regarding the letters ``\emph{j}'' and ``\emph{h}''---keys that are close on the keyboard. However, being the text in English, very simple techniques based on dictionaries can be applied to correct the error. If the text containing the erroneously inferred key was a password---typically with more entropy---dictionary based techniques would not work just as well. However, in these cases we can exploit movement speed, angle, and other possible features that define the particular way each user moves her finger or the stylus to type on the keyboard. It is very likely that this particularity impacts the key-inference accuracy of Hoover and that makes so that a specific couple of keys, like ``\emph{j}'' and ``\emph{h}'', tend to be interchanged. With this in mind, from the example above we can easily deduce that Jane's password for the \emph{paypal}  account is very likely to be \emph{jane1984}.

A deep analysis of the impact of user habits in Hoover's accuracy is out of the scope of this work. Nonetheless, it can give an idea on the strength and the pervasiveness of our attack. 

\subsection{User-biometrics Information} 
So far we have just discussed what an adversary can obtain by associating the user click streams stolen by Hoover to their semantics (e.g., text typed, messages exchanged with friends, and so on). But, the data collected by Hoover has a lot more potential than just this. In fact, it can be used to infer user biometric information and profiling regarding her interaction with the device. This is possible thanks to the timestamps of clicks collected by the Listener view. 

The Listener view in Hoover obtains timestamps each time a hover event is fired in the system. In particular, it obtains timestamps for events of the type \emph{touch down} (the user clicks) and \emph{touch up} (the user removes the input device from the screen). These timestamps allow Hoover to extract the following features: (i) the click duration (ii) the duration between two consecutive clicks, computed as the interval between two corresponding \emph{touch down} events (iii) hovering duration between two clicks, computed as the interval between a \emph{touch up} event and the next \emph{touch down} event. These features are the fundamentals for continuous authentication mechanisms based on biometrics of the user~ \cite{biometrics14, biometrics15}. In addition, the mechanisms proposed in~\cite{biometrics14, biometrics15} require a system level implementation, which can be tricky and add complexity to existing systems. To the best of our knowledge, Hoover is the first app-layer that offers a real opportunity for biometric-based authentication mechanisms. Hoover can continuously extract features from clicks to authenticate the device owner and differentiate her from another user, e.g., a robber who stole the device. 

While biometric-related information is a powerful means for authentication, the same information could also be misused in order to harm the user. For example, the authors in \cite{iCanBeYou} show how an adversary holding a large set of biometric-related information on a victim user can use it to train and bypass keystroke based biometric authentication systems. In this view, Hoover's potential to profile the way a user types could also be exploited to actually harm her in the future.

\section{Discussion and Countermeasures}
\label{sec:countermeasures}

The success of the attack we described relies on a combination of an unexpected use of hover technology and Alert Window views. Here we review possible countermeasures against this attack and we show that, what might seem straightforward fixes, either cannot protect against the attack, or severely impact the usability of the system or of the hover technology. 

\subsection{Limit Access to Hover Events} The attack presented in this paper exploits the information dispatched by the Android OS regarding hover events. In particular, the hover coordinates can be accessed by all views on the screen, even though they are created by a background app, like Hoover. A possible mitigation could be to limit the detection of hover events only to components (including views) generated by the application running in the foreground. In this way, despite the presence of the invisible overlay imposed by Hoover (running in the background), the attacker would not be able to track the trajectory of the movement while the user is typing. However, this restrictive solution could severely impact the usability of existing apps that use, in different ways, Alert Windows for a better user experience. An example is the ChatHead feature of the Facebook's Messenger application: If not enabled to capture hover events, this feature would be useless as it would not capture user clicks either. Recall that, a view either registers both clicks (touches) and hover events, or none of them at a single time.

Another possibility would be to decouple hover events from click events, and limit the first ones only to foreground activities. This solution would add complexity to the hover-handling components of the system and would require introducing and properly managing additional, more refined permissions. Asking users to manage (complex) permissions has been shown to be inadequate---most users tend to blindly agree to any permission requested by a new app they want to install~\cite{permissionsUser}. Not only users, but developers as well find already existing permissions too complex and tend to over-request permissions to ensure that applications function properly~\cite{developersPermissions}. Given this, introducing additional permissions does not seem like the right way to address this problem in an open system like the Android OS. Finally, another possibility is to eliminate or switch off the hover feature from the devices. Clearly this would introduce considerable issues regarding usability. Which is why Samsung devices partially allow this possibility for the fingers, while still keeping stylus hover in place.

\subsection{The Touch Filtering Specific in Android} 

Here we explain why the \emph{filterTouchesWhenObscured} mechanism~\cite{viewdev} cannot be used to thwart our attack.
First, we start off by shortly describing its functionality. The touch filtering is an existing Android OS specific that can be enabled or not for a given UI component, including a view. When enabled for a given view, all clicks (touch events) issued over areas of the view obscured by another service's window, will not get any touch events. That is, the view will never receive notifications from the system about those clicks. 

The touch filtering is disabled by default, but app developers can enable it for views and components of a given app by calling the \emph{setFilterTouchesWhenObscured(boolean)} or by setting the \emph{android:filterTouchesWhenObscured} layout attribute to true. 

If Hoover were to obstruct components during clicks, the touch filtering could have endangered its stealthiness---the component underneath which the click was intended for would not receive it, so the user would eventually be alerted. However, Hoover never obstructs screen areas during clicks. (Recall that the malicious overlay is created and destroyed in appropriate instants in time, so as to not interfere with user clicks, see Section~\ref{sec:our_approach}). So, even with the touch filtering enabled by default on every service and app, neither the accuracy, nor the stealthiness of Hoover are affected.

\subsection{Forbidding 0px views or the Activation of the \texttt{FLAG\_WATCH\_\-OUTSIDE\_\-TOUCH}}
Hoover uses a 0px view which listens for on-screen touch events and notifies the malware about the occurrence of a click so it can promptly activate the transparent overlay. Thus, forbidding the creation of 0px views by services seems like a simple fix to thwart the attack. However, the attacker can still overcome the issue by generating a tiny view and position it on screen so as to not cover UI components of the foreground app. For instance, it could be shown as a thin black bar on the bottom, thus  visually indistinguishable from the hardware border of the screen.

The Listener view also uses the \texttt{FLAG\_WATCH\_\-OUTSIDE\_\-TOUCH} to detect when the user clicks. It may seem that if this flag were disabled, the attack would be stopped. However, this same functionality can be achieved in two alternative ways without this flag: The first is to analyze information coming from sensors like gyroscope and accelerometer~\cite{tapprints2012}, accessible without permissions on Android. The second is to continuously monitor the information in the \emph{proc} folder related to the keyboard process, also accessible without permissions on Android~\cite{adbAttack}. Previous works have shown that both methodologies can be highly accurate in inferring click times~\cite{tapprints2012, adbAttack}. Not only that, this flag is used by many applications. For example, whenever a window has to react (e.g., to disappear) when the user clicks outside it. This is a very important flag, very commonly used, and it is hard to tell how many applications would stop working properly if this functionality were to be disabled. 

\subsection{Limiting Transparent Views to Legitimate or System Services}
This limitation would impede Hoover to exploit the transparent overlay. Nonetheless, it could be overcome by a more sophisticated attacker. For example, in the keyboard attack scenario, the overlay could be a non-transparent and exact copy of the keyboard image on the victim's phone. Note that the keyboard layout depends on the device specifications (type, model, screen size). This information can easily be obtained on Android through public APIs of the WindowManager Interface~\cite{windowManager}. The keyboard-like overlay would then operate just like the transparent one yet being equally undetectable by the user. A similar approach can be used to collect the clicks of a target app whose design and components are known to the attacker (e.g., a login page for a well-known app like Mobile Banking or Facebook, and so on).

\subsection{Inform the User About the Overlay, trusted paths} The idea is to make the views generated from a background service easily recognizable by the user by restricting their styling; e.g., imposing, at system level, a well-distinguishable frame-box or texture pattern, or both. In addition, the system should enforce that all overlay views adhere to this specific style, and forbid it for any other view type. However, countermeasures that add GUI components as trusted path~\cite{vigna15, windowGuard, guarDroid} to alert a user about a possible attack (security indicators) have not been shown to be effective. This is confirmed by the findings of an extensive user study in~\cite{vigna15}: even when the subjects were aware about the possibility of the attack and the countermeasure was in place, there were still 42\% of users that kept using their device normally. 

This kind of trusted paths can help against phishing attacks, when the user is interacting with a malicious app instead of the genuine app.
However, note that this is not the case for our attack where the malware always runs in the background and does not interact with the legit foreground app, the one the user is interacting with. Even if security indicators were to be shown on a view-based level rather than on an app-based one like in~\cite{vigna15}, note that the overlay in Hoover is not static. Rather, it is shown for very short time windows (70ms) successive to a click, when the user focus is probably not on the security indicator.

\subsection{Protecting Sensitive Views} The idea is to forbid that a particularly sensitive view or component generated by a service, like the keyboard during login sessions, or an install button of a new app, is overlaid by views of other services, including Alert Windows or Toasts. A possible implementation could be the following: Introduce an additional attribute of the view class that specifies whether a given instance of the class should or not be ``coverable''. When this attribute is set, the system enforces all other screen object overlapping with it to be ``pushed out'' the specific view's boundaries; e.g., in another area on the screen not covered by the view. Clearly, it would be a responsibility of the app-builder to carefully design her app and identify sensitive views that require the non-coverable attribute. In addition, these types of views should have a maximum size and not cover the whole screen. Otherwise, it would not be possible for other services, including system ones, to show Alert Windows in presence of a non-coverable view. This solution could mitigate attacks like ours and also others that rely on overlays even though in a different way, e.g., phishing or clickjacking. However, it would put a considerable burden on the app builders that will have to carefully classify UI components of their apps into coverable and non coverable, taking also into consideration possible usability clashes with views generated unexpectedly from other legit or system services like on screen message notifications, system Alert Windows, and so on. 

\section{Related Work}
\label{sec:related}

The main challenge to achieve the goal of inferring user input comes from a basic rule of Android: A click is directed to (and thus captured by) one app only. However, existing works have shown that malware can use various techniques to bypass this rule and infer user input (e.g., steal passwords). We can think of mobile application phishing~\cite{activityAttack} as a trivial case of input inference attacks, where the goal is to steal keyboard input (typically login credentials) of the phished application. Although effective when in place, a limitation of phishing attacks is their distribution through official app markets. Furthermore, and contrary to our techniques, phishing attacks need to be implemented separately for every phished app.

Hoover does not affect user experience. This makes it more robust and stealthy than UI redressing (e.g., clickjacking) attacks which also achieve input inference on mobile devices~\cite{clickjackingBlackhat12, embedded2013, taskJacking15, defenseSchemes2015, vigna15, clickjackingUSENIX14}. UI redressing techniques use overlay windows in a conceptually different manner with respect to our work. They typically cover a component of the application with an alert window or Toast message overlay~\cite{clickjackingBlackhat12, vigna15} that, when clicked, either redirects the user towards a malicious interface (e.g., a fake phishing login page), or intercepts the input of the user by obstructing the functionality of the victim application (e.g., an overlay over the whole on-screen keyboard). Such invasive attacks disrupt the normal user experience: The victim application never gets the necessary input, which can alarm the users.

An alternative approach is  to infer user input in a system-wide manner through side-channel data obtained from various sensors present on the mobile platform~\cite{soundComber2011, taplogger2012, acCompliance2012, tapprints2012, placeRaider2013, starbug14, gyroUSENIX14, Narain2014, nearbykeyboard2011}, like accelerometers and gyroscopes. Reading such sensor data commonly requires no special permissions. However, such sensors provide signals of low precision which depend on environmental conditions (e.g., the gyroscope of a user that is typing on a bus in movement). So, the derived input position from these side-channels is often not accurate enough to differentiate, e.g., which keys of a full on-screen keyboard were pressed. Conversely, Hoover proves to work with high accuracy and even infer all user keystrokes with 2px error in case of stylus. Differently, the microphone based keystroke inference~\cite{Narain2014} works well only when the user is typing in portrait mode. In addition, its accuracy depends on the level of noise in the environment. 

Contrary to related works, our attack does not restrict the attacker to a given type of click-based input (e.g., keyboard input inference only), but targets all types of user clicks. It does not need to be re-implemented for every target app, like phishing and UI redressing, as it works system-wide. 

\section{Conclusion and Future Work}
\label{sec:discussion}

In this work we proposed a novel type of user input inference attack. We implemented \emph{Hoover}, a proof-of-concept malware that records user clicks performed by either finger or stylus as input device, on devices that support the hover technology. In contrast to prior works, Hoover records all user clicks with high precision  and granularity (at the level of click positions on screen). The attack is not tailored to any given application, operates system-wide, and is transparent to the user: It does not obstruct the normal user interaction with the device in any way. 

In this work, we did not distinguish between specific fingers as input methods. However, our initial experiments pointed out that training per-finger models increases the attack accuracy. Employing techniques for detecting which finger the user is using~\cite{Goel:2012}, and using the correct finger model could potentially improve the accuracy of our attack. We leave this as future work.

\bibliographystyle{ACM-Reference-Format}
\bibliography{bibliography}

\end{document}